 \documentclass[aps,prb,twocolumn,groupedaddress,showpacs]{revtex4}

\usepackage{graphicx}
\usepackage{dcolumn}
\usepackage{bm}

\begin{document}

\newcommand{\Vec}[1]{\bm{#1}}
\newcommand{\I}[0]{\text{i}}


\title{ Application of the Density Matrix Renormalization Group in
momentum space }
\author {Satoshi Nishimoto}
\author {Eric Jeckelmann}
\author {Florian Gebhard}
\affiliation{Fachbereich Physik, Philipps--Universit\"at Marburg, 
D--35032 Marburg, Germany}
\author{Reinhard M.~Noack}
\affiliation{Institut f\"ur Physik, Johannes--Gutenberg--Universit\"at
Mainz, D--55099 Mainz, Germany 
}
\date{\today}

\begin{abstract} 

We investigate the application of the Density Matrix Renormalization
Group (DMRG) to the Hubbard model in momentum--space.
We treat the one--dimensional models with dispersion relations
corresponding to nearest--neighbor hopping and $1/r$ hopping
and the two--dimensional model with isotropic nearest--neighbor hopping.
By comparing with the exact solutions for both one--dimensional models
and with exact diagonalization in two dimensions, we first investigate the
convergence of the ground--state energy.
We find variational convergence of the energy with the number of
states kept for all models and parameter sets.
In contrast to the real--space algorithm, the accuracy becomes rapidly
worse with increasing interaction and is 
not significantly better at half filling.
We compare the results for different dispersion relations at fixed
interaction strength over bandwidth and find that
extending the range of the hopping in one dimension has
little effect, but that changing the dimensionality from one to two
leads to lower accuracy at weak to moderate interaction strength.
In the one--dimensional models at half--filling, 
we also investigate the behavior of the single--particle gap, the
dispersion of spinon excitations, and the momentum distribution
function.
For the single--particle gap, we find that proper extrapolation in the
number of states kept is important.
For the spinon dispersion, we find that good agreement with the exact
forms can be achieved at weak coupling if the large momentum--dependent
finite--size effects are taken into account for nearest--neighbor
hopping.
For the momentum distribution, we compare with various weak--coupling and
strong--coupling approximations and discuss the importance of
finite--size effects as well as the accuracy of the DMRG.

\end{abstract}


\maketitle

\section{Introduction}

Many renormalization schemes are carried out in momentum
space and involve integrating out degrees of freedom using a momentum
cutoff. 
For example, Wilson's numerical Renormalization Group (RG)~\cite{wilson} 
implements
this program using a mapping of momentum shells to an
effective lattice model.
The renormalization process is carried out by successive numerical
diagonalization of a finite system and energetic truncation of the
Hilbert space.
While this lattice model corresponds to successive momenta or,
equivalently, energy scales, its form is similar to that of a strongly
correlated lattice model.

Attempts at applying a real--space version of the Wilson procedure to
short--range quantum 
lattice models such as the Heisenberg or the Hubbard model were not
successful, however, because successive lattice points do not
correspond to different energy scales.
The Density Matrix Renormalization Group (DMRG)~\cite{steve1,book}
overcomes these
limitations by carrying out the
renormalization on a subsystem.
The truncated basis is formed by projecting the state of the entire
system onto the subsystem using the reduced density matrix rather than
selecting states energetically.
This method has been very successful at treating low--dimensional
quantum lattice models with open boundary conditions and short--range
couplings. 
However, for longer--range off--diagonal interactions, higher
dimensional systems or lattices with periodic boundary conditions,
this real--space formulation of the DMRG is much less successful.
In addition, it loses sight of an energy or momentum--based
classification of the relevant degrees of freedom.

A potential way of overcoming this limitation for itinerant electron
systems is to apply the DMRG
ideas to the momentum--space formulation of the Hamiltonian.
This approach has a number of potential advantages over the real--space
approach.
First, since the single--particle basis in momentum space is explicitly
translationally invariant, momentum is a conserved quantum
number.
Use of this momentum quantum number reduces the
size of the Hilbert space in the diagonalization.
Second, momentum--dependent quantities such as the momentum
distribution or the dispersion of excitations can be directly calculated.
Third, the kinetic energy term is diagonal so that
varying the dispersion by, for example, changing the range
of the hopping, is easy to do.

Attempts to formulate a numerical renormalization group procedure for
quantum lattice systems in momentum space~\cite{steve2} predating the
DMRG were not particularly successful~--~this was one of White's
motivations for turning to real space and formulating the DMRG.
Shortly after the development of the DMRG in real--space, White
attempted to use DMRG methods on the momentum--space formulation of
the Hubbard model.
He calculated the ground--state energy in
one and two dimensions at intermediate couplings, but found 
that the energies obtained were not significantly better than
those obtained by other variational methods.~\cite{steveprivate}

Independently, Xiang developed a similar technique and applied it to
the Hubbard model in one and two dimensions.~\cite{xiang}
In this work, Xiang outlined an efficient implementation of the DMRG
in momentum space.
He developed a factorization of the Hubbard interaction that reduces the
number of term from $N^3$, where $N$ is the number of single--particle
Bloch wavefunctions in the lattice, to $6N$.
He also pointed out some features of the algorithm that need to be
carefully considered in momentum space:
Since the interaction is highly non--local, there is no natural
ordering of the single--particle states; the choice of the ordering
can, however, have an effect on the performance of the DMRG
algorithm.
In addition, there is no well--defined infinite--system algorithm, so that
care must be taken in how the lattice is built up initially.
Care must also be taken in this initialization procedure so that
states with a sufficient spread in momentum quantum numbers are kept.
One possible outcome of an inadequate initialization procedure is
convergence to a state other than the true ground state.

Xiang investigated the performance of the algorithm for various
interaction strengths for the one--dimensional Hubbard model with 16
sites at half filling, and for the two--dimensional model on
system sizes ranging from $4\times 4$ to $12\times 12$ for various
band fillings.
He found that the convergence of the method depends strongly on the
interaction strength, $U$.
The method is exact for $U=0$ since the
Hamiltonian is diagonal and the convergence becomes rapidly worse
with increasing $U$.
In one dimension, he compared with real--space DMRG calculations and
found that the error in the ground--state energy was higher than the
real--space calculation for both weak and intermediate interaction
strengths ($U/t=1$ and 4, with $t$ the hopping matrix element), with
5\% error for $U/t=4$.
In two dimensions, he compared with exact diagonalization for a
$4\times 4$ system, cluster diagonalization on a $6\times 6$ system,
and quantum Monte Carlo and stochastic diagonalization calculations on
$4\times 4$, $6\times 6$ and $8\times 8$ systems.
The relative errors increased rapidly with $U$ for the $4\times 4$
system.
The variational energies were comparable to those of the stochastic
diagonalization and quantum Monte Carlo methods (for which the energy
is non--variational) for larger system sizes.
The variational bounds for the energy were slightly higher than
stochastic diagonalization for the $6\times 6$ lattice and slightly
lower for the $8\times 8$ lattice.
In comparing the performance in one and two dimensions, Xiang pointed
out that the accuracy for given number of states kept,
$U/t$, and band filling for 16 site systems was better in two
dimensions than in one, leading him to speculate that the
momentum--space method becomes more accurate as the dimensionality is
increased.

Our purpose in this work is to explore more fully both the
convergence properties and the application of the momentum--space
formulation of the DMRG to the Hubbard model.
In one dimension, we take advantage of the existence of exact
solutions for two choices of the dispersion, corresponding to
nearest--neighbor hopping and $1/r$ hopping, to systematically
investigate the dependence of the convergence of the ground--state
energy on interaction strength,
band filling, and number of density-matrix eigenstates kept, $m$.
We investigate the regularity of the convergence with $m$ and discuss
schemes to extrapolate in $m$ in order to obtain more accurate
energies.
We reexamine the relative convergence for the one and
two--dimensional models with a view to understanding the utility of
the momentum--space DMRG for higher--dimensional systems.

While the ground--state energy is useful for determining
variational convergence, it does not directly provide much useful
information about the physical behavior of the system.
We therefore investigate some physically useful quantities, 
the quasiparticle
gap, the momentum distribution and the dispersion of spinon
excitations, for the one--dimensional models and compare to exactly
known results and perturbative approximations, where appropriate.
Our calculations of the momentum distribution and the spinon dispersion
for the $1/r$--hopping model are, to our knowledge, 
the first
independent numerical calculations of these quantities.

The layout of the remainder of this paper is as follows: 
In section \ref{sec:model} we
discuss the model systems and their basic properties. 
Our DMRG method is described in section~\ref{sec:method}.
The convergence and accuracy of the momentum--space DMRG are discussed
in section~\ref{sec:convergence}. 
We study the dispersion of spinon excitations
and the momentum density distribution of one--dimensional Hubbard models
in sections~\ref{sec:spinon} and \ref{sec:momdens}, respectively.
We discuss the prospects for momentum--space DMRG in the
final section.

\section{Model}
\label{sec:model}

The Hubbard model~\cite{hubbard} is defined in a general form by 
the Hamiltonian
\begin{equation}
H = \sum_{i,j,\sigma} t_{ij} c_{i \sigma}^\dagger c_{j \sigma} 
+ U \sum_i n_{i \uparrow} n_{i \downarrow}
\label{eq:ham_r}
\end{equation}
where $c_{i \sigma}^\dagger$ ($c_{i \sigma}$) creates (annihilates) an 
electron with spin $\sigma$ in the Wannier state
on lattice site $i$ with position $\Vec{r}_i$, 
$n_{i\sigma}=c_{i \sigma}^\dagger c_{i \sigma}$ denotes the 
particle number operator on site $i$, $t_{ij}=t(\Vec{r}_i-\Vec{r}_j)$ 
is the transfer integral between site $i$ and $j$, 
and $U$ is the energy cost due to the Coulomb repulsion of two 
electrons on the same site.  
In this paper, all energies are measured in units of $t=1$.  

Using the relation (Fourier transformation) 
between Wannier states at site $i$ and Bloch states with 
momentum~$\Vec{k}$
\begin{equation}
c_{\Vec{k} \sigma}^\dagger =  \frac{1}{\sqrt{N}}
\sum_j e^{\I \Vec{k} \cdot \Vec{r}_j} c_{j \sigma}^\dagger  \, ,
\end{equation}
where $N$ is the number of lattice sites,
the Hubbard Hamiltonian with translationally invariant
transfer integrals is written in momentum space as
\begin{equation}
H = \sum_{\Vec{k}, \sigma} \varepsilon (\Vec{k}) 
n_{\Vec{k} \sigma} 
+ \frac{U}{N} \sum_{\bm {p,k,q}} 
c_{\Vec{p}-\Vec{q} \uparrow}^\dagger 
c_{\Vec{k}+\Vec{q} \downarrow}^\dagger c_{\Vec{k} \downarrow} 
c_{\Vec{p} \uparrow}   \, ,
\label{eq:ham_k}
\end{equation}
where  $n_{\Vec{k} \sigma} 
= c_{\Vec{k} \sigma}^\dagger c_{\Vec{k} \sigma}$ and 
\begin{equation}
\varepsilon (\Vec{k})= \sum_j e^{-\I \Vec{k} \cdot \Vec{r}_j} 
\ t(\Vec{r}_j)
\end{equation}
is the energy dispersion of the electrons.

The kinetic energy of Eq.\ (\ref{eq:ham_k}) consists only of diagonal terms
with dispersion $\varepsilon (\Vec{k})$,
so that the momentum--space DMRG method is trivially exact
for $U=0$.
Moreover, it can be easily applied to different non--interacting 
dispersions $\varepsilon (\Vec{k})$ corresponding to different lattice 
geometries and hopping ranges.  
In this paper, we apply the momentum--space DMRG to the following three 
different models (here and in what follows, we take the lattice
constant to be unity and $N$ to be even):

(i)~The one--dimensional Hubbard chain with nearest--neighbor hopping
amplitude $t_{j+1,j}=-t e^{\I\phi}$. 
The dispersion relation is given by
\begin{equation}
\varepsilon (k) = -2t  \cos (k-\phi)
\label{eq:1DHubbard}
\end{equation}
with $k=2 \pi n/N$ and $n=-N/2+1, \cdots, N/2$.
The bandwidth is $W=4t$.
Here a flux $N \phi$, measured in units of the flux 
quantum $\phi_0=hc/e$ and equivalent to a twisted boundary condition, 
is threaded through the system.~\cite{kohn,shastry-suth} 
It enables us to calculate the ground--state energy
as a function of flux $\phi$. 
We will use nonzero values of $\phi$ later to interpolate
momentum--dependent quantities (e.g.~the momentum distribution
function) to arbitrary values of the momentum on a finite system.

(ii)~The one--dimensional Hubbard chain with long--range hopping amplitude
\begin{equation}
t_{lm} = (-\I t) \frac{(-1)^{l-m}}{d(l-m)}
\end{equation}
with
\begin{equation}
d(l-m)=\frac{N}{\pi}\sin 
\left[ \frac{\pi(l-m)}{N}\right] \; .
\end{equation}
Since $d(l-m)$ is antisymmetric under the permutation of $l$ and $m$, 
the hopping matrix 
element has to be purely imaginary to guarantee that $t_{lm} = t_{ml}^*$.
In the thermodynamic limit ($N \to \infty$), 
the hopping decays proportionally 
to the inverse of the distance $r = |l-m|$
(``$1/r$--Hubbard model'').  
The dispersion relation is 
\textsl{linear} and is given by
\begin{equation}
\varepsilon (k) = tk
\end{equation}
with $k = (2n-1) \pi / N$ and $n=-N/2+1, \cdots, N/2$ where
antiperiodic boundary conditions are chosen.  
The bandwidth is $W=2 \pi t$.

(iii)~The two--dimensional Hubbard square lattice with nearest--neighbor
hopping amplitude $-t$.  
The dispersion 
relation is given by
\begin{equation}
\varepsilon (\Vec{k}) = -2t  (\cos k_x + \cos k_y)
\end{equation}
with $\Vec{k} = (2 \pi n_{x} / L, 2 \pi n_{y}/L)$ where
$n_x, n_y = -L/2+1, \cdots, L/2$ and 
$N = L^2$.  
The bandwidth is~$W=8t$.

The one--dimensional Hubbard model~(i) is exactly solvable
via the Bethe Ansatz~\cite{liebwu}
and can easily be studied using the real--space DMRG.~\cite{reinhard} 
Comparison with the exact solutions and the real--space 
method will provide
an opportunity to test the performance of the momentum--space DMRG.  
The $1/r$--Hubbard model~(ii) is also exactly solvable,~\cite{florian} but
it is difficult to investigate with the real--space DMRG because the
hopping is long--range and imaginary.
For this model, the advantage of the momentum--space approach
is clear: one need only change the real, diagonal dispersion
$\varepsilon (k)$.
We will therefore use this model to investigate how a substantial change in
the range of hopping affects the momentum--space algorithm.
The Hubbard model on a two--dimensional square lattice (iii) will be
used to investigate and compare the effects of dimensionality on the
momentum--space method and on the real--space method.

\section{DMRG in momentum space}
\label{sec:method}

In principle, the usual DMRG~\cite{steve1,book}
can be applied directly to
the momentum--space representation of the Hubbard model (\ref{eq:ham_k}).
In the momentum--space approach each Bloch function with momentum 
$k$ and spin $\sigma$ corresponds to a lattice site. 
To perform the calculations presented in this paper,
we have adapted a program originally written by White.\cite{steveprivate}
This program predates (and thus does not use)
some recent developments which can greatly improve the performance of
the DMRG
such as the wavefunction transformation~\cite{steve3,book},
the use of composite operators,~\cite{xiang}
and of non--abelian symmetries.~\cite{McCulloch}
Nevertheless, this program is highly optimized and allows
us to carry out DMRG calculations
keeping up to $m=4000$ density--matrix eigenstates
on a workstation with 1GB of memory.
Below and in the next section
we discuss features of the momentum space
DMRG which differ from the usual real space DMRG. 

The summation in the second term of Eq.~(\ref{eq:ham_k}) runs
over $N^3$ products of operators. 
A straightforward implementation of 
the DMRG algorithm requires calculating and 
keeping  track of ${\cal O}(N^3)$ matrices representing
the different products of operators.  
This represents a significant increase 
compared to the real space-approach which requires
only a constant number of operators for the one--dimensional and order
${\cal O}(L = \sqrt{N})$ matrices
for the two--dimensional Hubbard model~(\ref{eq:ham_r}),
respectively.
Xiang~\cite{xiang} has shown that is it possible
to define so--called composite operators and thus
reduce the number of operators which need to be kept to $6N$.
In White's program, internal sums over blocks are carried out to
reduce the number of operators to ${\cal O}(N^2)$ rather than 
${\cal O}(N)$, 
and an efficient representation of operators with small sparse matrices
is used.

We explicitly use the conservation of the particle number
$N_{\text{e}} =  \sum_{\Vec{k}, \sigma} n_{\Vec{k} \sigma}$,
of the $z$-component of the total spin $S_z = 
(1/2)\sum_{\Vec{k}, \sigma} \sigma \; n_{\Vec{k} \sigma}$, and
of the total momentum 
\begin{equation}
\Vec{K} = \sum_{\Vec{k}, \sigma } \; \Vec{k} \;
n_{\Vec{k} \sigma} \quad  
\text{mod} \quad 2\pi  \; .
\end{equation}
Momentum symmetry reduces the size of the effective Hilbert space by
about a factor $N$ 
and allows us to decompose the matrix representations
of operators into several smaller matrices.
Therefore, the dimension of the effective Hilbert space
for a given number $m$ of density matrix eigenstates
is smaller in the momentum--space approach
than in the real space approach.
This should be kept in mind when comparing results
obtained with both approaches: In general, $m$~can be made larger for a
given amount of computational effort in momentum space.

In a one--dimensional system in real space, there is a natural 
ordering of the lattice sites. 
In two dimensions, there is some choice in the
ordering,~\cite{reinhard,xiang2} but reasonable
choices yield similar results.~\cite{peschel}
In momentum space it is not a priori clear how one should arrange
the sites in the lattice.\cite{xiang}
Thus, we have tested several possibilities
including random ordering.
We have found that the order of sites should be carefully
chosen in the momentum--space approach --~the rate of 
convergence and the accuracy 
strongly depends on the site order. 
Fundamentally, it seems that Bloch states which are
strongly scattered by the Hubbard term in (\ref{eq:ham_k})
should be arranged to be as close together as possible. 
For the one--dimensional and two--dimensional Hubbard models
with nearest--neighbor hopping
we use an energetic ordering in which the sites
are arranged according to $|\varepsilon(\Vec{k})-\varepsilon_{\text{F}}|$,
where $\varepsilon_{\text{F}}$ denotes the Fermi energy in the
non--interacting case ($U=0$). 
For the $1/r$--Hubbard model, the Fermi energy has no particular
relevance for low--energy scattering processes and 
we have found that ordering the sites according to $\varepsilon(k)$
works best.

We use the finite--system DMRG algorithm and perform  
several sweeps through the lattice until the ground--state
energy converges as in the real--space approach. 
Wilson's numerical RG method is used instead of the infinite
system algorithm to build up the lattice during the initial iteration. 
For the next iterations we apply the usual blocking scheme
with a superblock made of two sites and two blocks with at most
$m$ states each.
In Ref.~\onlinecite{xiang} the superblock was built
using two blocks and a single site.
According to Xiang, this single--site approach is faster than the usual
blocking scheme.
As discussed in Ref.~\onlinecite{steve1}, however,
the single--site blocking scheme is not a robust method unless
several states are targeted.
As we always target a single state in our calculations
[the ground state for some quantum numbers 
$(N_{\text{e}},S_z,\Vec{K})$],
we use the two--site blocking scheme. 

We have observed that the DMRG has difficulty finding  
the ground state when the interaction $U$ is not weak 
or for some particular choices of the quantum number
$(N_{\text{e}},S_z,\Vec{K})$. 
The DMRG sometimes seems to converge first
to a state other than the ground state and only converges to the true
ground state after many sweeps or after the number $m$
of density--matrix eigenstates is increased.
This behavior is marked by a rapid drop in the energy after a
relatively large number of sweeps or at a high value of $m$.
A similar problem has been reported with the real--space
DMRG applied to two--dimensional fermion
systems.~\cite{steve4,bonca}
Therefore, one should not rely on DMRG results obtained for
a fixed number $m$ of density--matrix eigenstates kept or 
a fixed number of sweeps, but one should investigate the behavior
of the DMRG as a function of  $m$ and of the number of sweeps.

\section{Convergence and accuracy of momentum space DMRG}
\label{sec:convergence}

In this section we discuss the convergence and accuracy of 
the momentum--space DMRG method applied to the Hubbard
model~(\ref{eq:ham_k}). 
We have also applied the real--space DMRG method to the
real--space representation~(\ref{eq:ham_r}) of the Hubbard
Hamiltonian with periodic boundary conditions. 
This allows us to make comparisons of
both DMRG methods in order to illustrate both their differences
and their common features.  
Our real--space DMRG program uses more advanced techniques
and is better optimized
than our momentum--space DMRG program. 
Therefore, we have chosen to present no comparison of computer CPU 
time and memory usage in this paper as they would be meaningless.

As a measure of the DMRG precision we use the error in the
ground--state energy per site
\begin{equation}
\Delta E(m) =  \frac{E_{\text{DMRG}}(m) - E_{\text{exact}}}{N t} \; ,
\end{equation}
where $E_{\text{exact}}$ is the exact ground--state energy
[for particular quantum numbers ($N_{\text{e}}, S_z,\Vec{K}$)]
and $E_{\text{DMRG}}(m)$ is the corresponding DMRG energy obtained
with $m$ density--matrix eigenstates kept.
The exact results $E_{\text{exact}}$ are calculated using the
Bethe Ansatz for the one--dimensional Hubbard model~\cite{liebwu}
and are computed numerically using exact diagonalization
techniques for the two--dimensional Hubbard model.~\cite{fano}  
For the $1/r$--Hubbard model they are derived from a conjectured
effective Hamiltonian.~\cite{florian}
Consistency of the DMRG energies with the spectrum obtained from the
effective Hamiltonian in turn confirms the conjecture.

\begin{figure*}
\includegraphics[width=7.5cm]{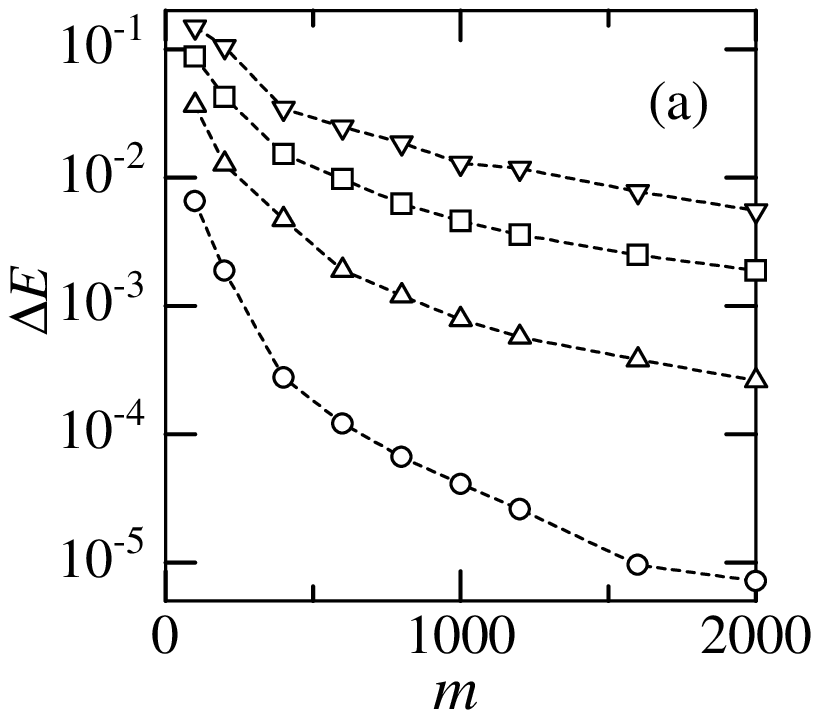}\hspace*{0.5cm}
\includegraphics[width=7.5cm]{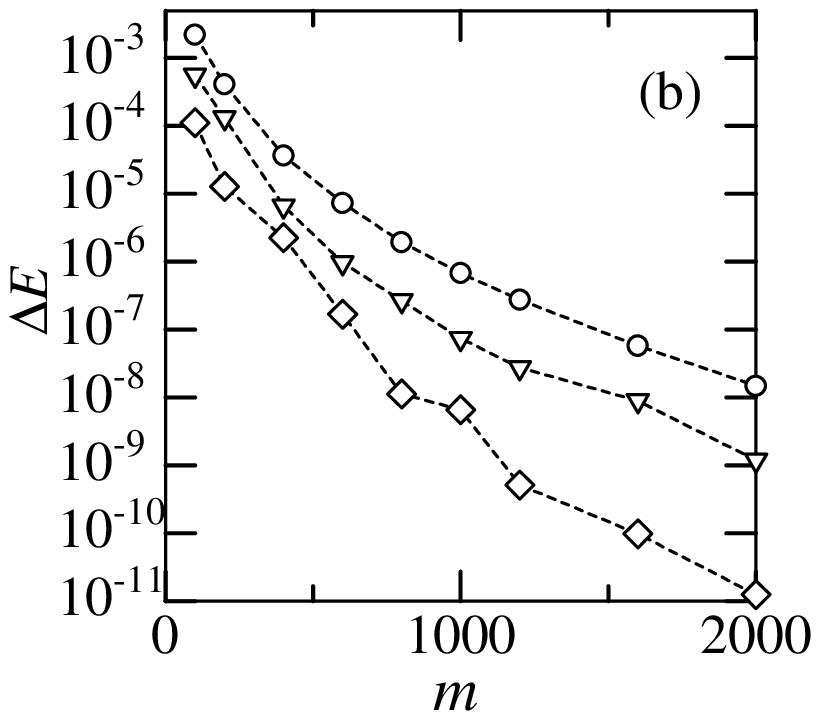}
\caption{DMRG error in the ground--state energy of the half--filled
one--dimensional Hubbard model with
16 sites and periodic boundary conditions as
a function of the number $m$ of density--matrix eigenstates kept in (a) the
momentum--space approach for $U/t=1$ (circles), 2 (triangles), 
3 (squares), 4 (reverse--triangles), and (b) the real--space approach
for $U/t=1$ (circles), 4 (reverse--triangles), 8 (diamonds).  }
\label{fig1} 
\end{figure*}

Figures~\ref{fig1}(a) and (b) show the ground--state energy 
error $\Delta E$ of the one--dimensional Hubbard model
as a function of the number of density--matrix eigenstates $m$
for several values of $U/t$.
These results have been obtained on 16--site lattices with
periodic boundary conditions at half filling using
the momentum--space [Fig.~\ref{fig1}(a)] and the real--space
[see Fig.~\ref{fig1}(b)] approaches.
The error $\Delta E$ of the momentum--space DMRG method
clearly increases with $U/t$.
The error in the real--space DMRG increases with \textsl{decreasing}
$U/t$ for this half--filled system.
In the momentum--space DMRG, the procedure becomes exact when the
off--diagonal interaction terms vanish (i.e., at $U=0$) and should be
more accurate when they are small --~it is a weak--coupling
method.
In contrast, the real--space representation becomes exact 
(i.e., local) when $t\rightarrow 0$.
It is important to note that this is not equivalent to the large $U/t$
limit, in which the real--space DMRG does \textsl{not} become exact.
The increase in accuracy with~$U/t$ shown here is specific to the
half--filled insulator, in which the charge degrees of freedom become
increasingly localized with increasing~$U/t$.
In fact, in the one--dimensional system away from half band--filling, 
there is \textsl{very little} dependence
of the convergence on~$U/t$ (for open boundary conditions).~\cite{kneer}
In both approaches, $\Delta E(m)$ does not
decrease exponentially as $m$ increases, contrary to 
the behavior often reported for real--space DMRG calculations
on one--dimensional systems with open boundary conditions.
In Figs.~\ref{fig1}(a) and~(b)
we also see that the errors of the real--space approach
are smaller and decrease more rapidly with $m$ than the errors
of the momentum--space approach.

\begin{figure}
\includegraphics[width=8cm]{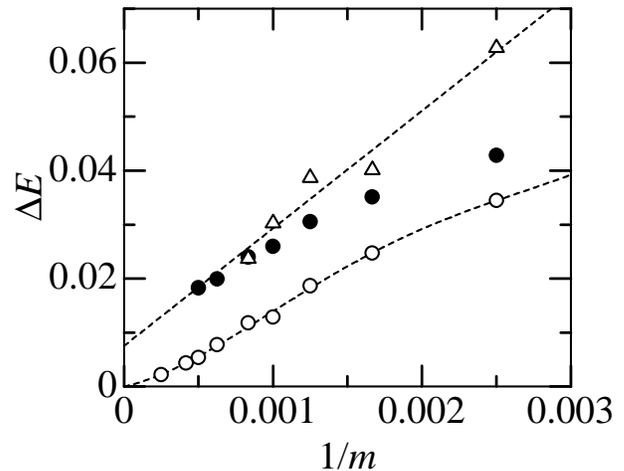}
\caption{DMRG error in the ground--state energy as a function 
of the number $m$ of density--matrix eigenstates kept for the 16--site
one--dimensional Hubbard model at half filling with $U/t=4$, including
our results for momentum $K=\pi$ (open circles) and $K=0$ 
(filled circles) and Xiang's results~\protect\cite{xiang} for unspecified
momentum (triangles).
The dashed lines are guides to the eye.  }
\label{fig2} 
\end{figure}

\begin{figure*}
\includegraphics[width=7.5cm]{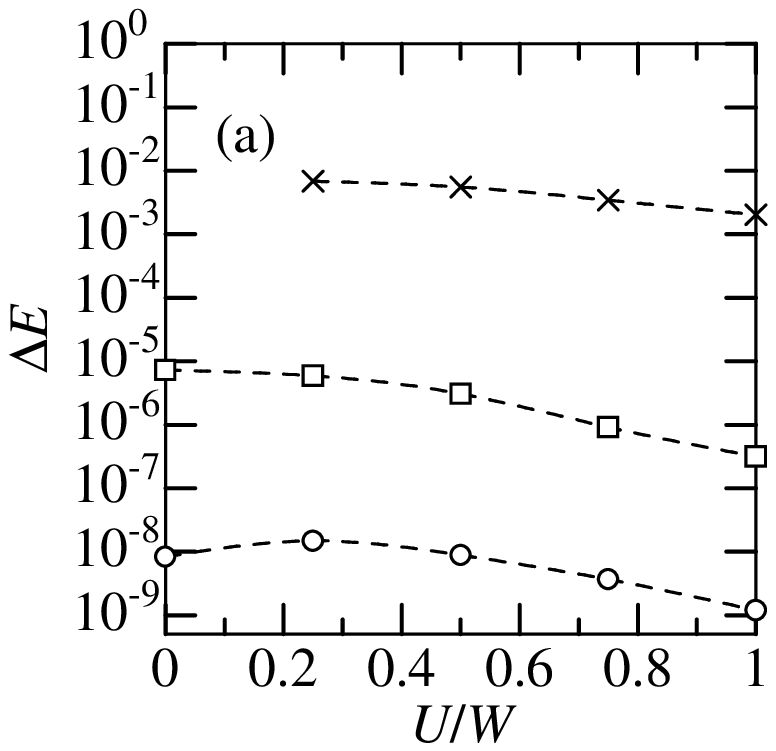}\hspace*{0.5cm}
\includegraphics[width=7.5cm]{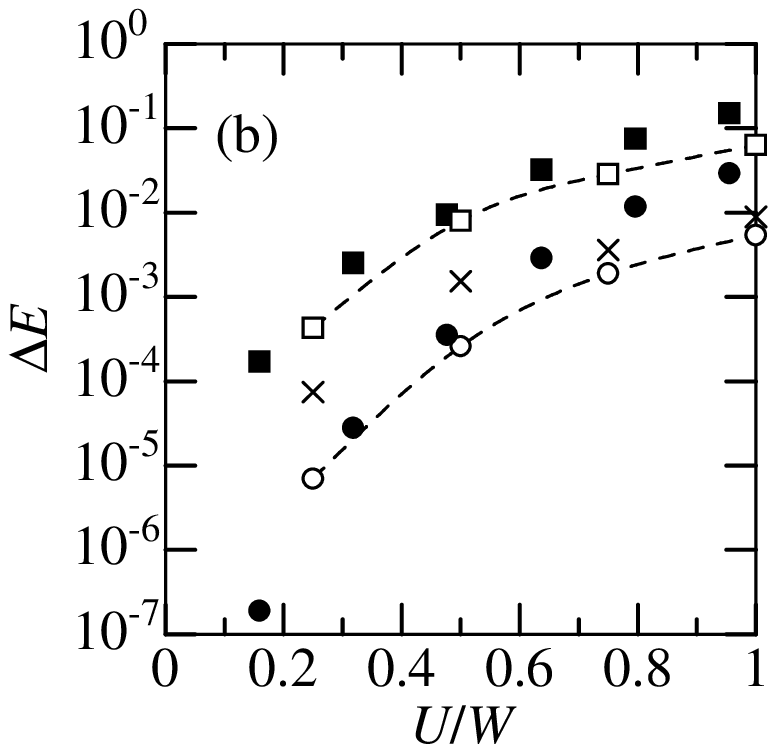}
\caption{DMRG errors in the ground--state energy at half filling
as a function of the interaction strength $U/W$
calculated with $m=2000$ density--matrix eigenstates.  
The dashed lines are guides to the eye.  
(a) Real--space DMRG results for the 
Hubbard model on a one--dimensional lattice with 16 sites (circles) 
and 32 sites (squares), and on a two--dimensional
$4 \times 4$ lattice (crosses).  
(b) Momentum--space DMRG results for the one--dimensional 
Hubbard model with 16 sites (open circles) and 32 sites 
(open squares), the $1/r$--Hubbard model with 16 sites (filled circles) 
and 32 sites (filled squares), and the two--dimensional Hubbard model
on a $4 \times 4$ lattice (crosses).  }
\label{fig3} 
\end{figure*}

In Xiang's work,~\cite{xiang} the systematic convergence of the
momentum--space method seems to break down when the interaction
strength $U$ approaches the bandwidth $W$. 
In particular, for the half--filled Hubbard model on a one--dimensional
16--site lattice with $U/t=4$, the ground--state energy obtained from
the DMRG, shown in Figure~\ref{fig2}, does
not seem to converge smoothly toward the exact ground--state energy as
the number $m$ of retained density--matrix eigenstates is increased 
--~the results are oscillatory and hard to extrapolate.
While the origin of this irregular convergence in Xiang's data is
unclear, one factor that is essential to consider is that the
momentum of the ground state for a half--filled ring with 16 sites and
periodic boundary conditions is $K=\pi$ because it is an 
open--shell configuration.
Our results for $K=\pi$ do converge smoothly to the exact solution as a 
function of $1/m$, as seen in Fig.~\ref{fig2}.
The lowest--energy state with $K=0$, also shown in the
figure, lies closer to Xiang's results for the larger values of $m$,
but converges smoothly to an energy that is clearly higher than the
ground state. 
The deviation of Xiang's result could either be due to his ground state
having a different momentum or due to convergence of the
the DMRG to a state other than the ground state, as
discussed in the previous section.

For weaker interaction, $U/t=1$, Xiang's results
converge to the ground--state energy for $K=\pi$, but lie
significantly above our own results for the same number
of density--matrix eigenstates kept.  For instance,
the DMRG error in the ground--state energy $\Delta E(m)$ 
for $m=1200$ is one order of magnitude smaller in our calculations
than the value reported by Xiang. 
This also suggests an incomplete
convergence of DMRG in Xiang's calculations even for weak coupling.
However, it should also be kept in mind that in his work Xiang used
a different superblock structure with a single site between two blocks. 
As a consequence, the dimension of the effective Hilbert
space for a given number of states $m$ is smaller in Xiang's 
calculations than in our calculations. 
This difference
could be responsible for the better accuracy of our results
in the weak coupling limit ($U/t=1$) but cannot explain the
discrepancy observed for intermediate coupling ($U/t=4$).

In our calculations we have often observed that the DMRG energy
initially seems to converge towards a value larger than the exact
ground--state energy.
Upon further increasing the number of states $m$ or the number of sweeps,
it then converges to the exact result.
In all the cases we present here, the momentum--space DMRG yields
energies that \textsl{do} ultimately converge to the exact result, even
for very large interaction strength $U$, although the rate of this
convergence and the accuracy deteriorate rapidly as $U$ increases.

\subsection{Dependence on model parameters}

In this section, we discuss the dependence of the accuracy of the
energy of momentum--space DMRG
on the model parameters: the single--electron
dispersion $\varepsilon(\Vec{k})$,
the interaction strength $U/W$, the lattice dimensionality, and the
band filling.
Figures~\ref{fig3}(a) and (b)
show the ground--state energy error $\Delta E$ 
as a function of the interaction strength $U/W$ 
for a fixed number $m=2000$ of density--matrix eigenstates.
We again see that errors decrease in the real--space approach 
[Fig.~\ref{fig3}(a)] but increase in the momentum--space approach 
[Fig.~\ref{fig3}(b)] for increasing $U/W$.
In both cases errors increase with the system size
and are larger in two dimensions than in one dimension
for the same number of lattice sites.
However, the dependence on
system size and dimensionality of the momentum--space approach is
clearly weaker than in the real--space approach.
The lower precision of the DMRG in higher dimension
is easily understood as a consequence of increasing
off--diagonal coupling in the real--space approach.
A possible explanation for the slight decrease in accuracy with
dimension in the momentum--space representation is that although
the single--electron dispersion $\varepsilon(\Vec{k})$
remains diagonal for any dimension,
a larger proportion of single--electron states are close 
to the Fermi surface and are thus strongly scattered.
In Figs.~\ref{fig3}(a) and (b)
one also sees that the precision of the real--space approach is 
generally better than that of the momentum--space approach
in the one--dimensional Hubbard model. 
The latter becomes more accurate than the former for $U/t\alt 1$ only. 
In two dimension, however, the real--space approach performs
very poorly for periodic boundary conditions. 
The momentum--space approach yields better results for $U \alt W = 8t$.
We finally note that DMRG errors seem to be affected only a small
amount by the form of the dispersion $\varepsilon(\Vec{k})$ in the
momentum--space approach. 
In the real--space approach, however, changing the single--electron 
dispersion by introducing longer--range hopping lowers the DMRG
performance very rapidly.
In summary, we find that the momentum--space approach  
is superior to the real--space approach for applications
to translationally invariant systems with weak to intermediate
Coulomb interactions in two dimensions or on one--dimensional lattices
with long--range hopping. 

\begin{figure}
\includegraphics[width=8cm]{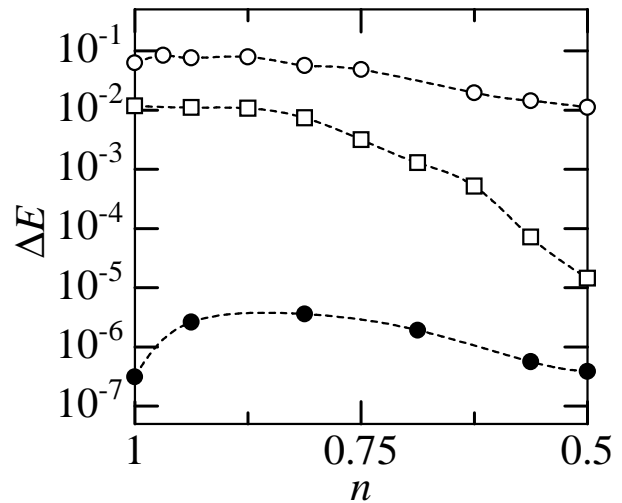}
\caption{DMRG error in the ground--state energy as function of
band filling $n$ in the one--dimensional Hubbard model
for $U/t = 4$ on lattices with 16 sites (squares)
and 32 sites (circles). 
Open symbols represent the momentum--space
DMRG results and filled symbols for the real--space DMRG results.
Real--space DMRG errors for 16 sites are smaller than $10^{-7}$
and not shown.
The number of density--matrix states is $m=1200$ and $m=2000$
for the 16--site and 32--site system, respectively.}
\label{fig4} 
\end{figure}

Let us now consider the effects of the band filling.
In Fig.~\ref{fig4} we show the error $\Delta E$ in the ground--state energy
as a function of band filling $n=N_{\text{e}}/N$ 
for the one--dimensional Hubbard model at $U=4t$. 
In the momentum--space approach, the accuracy is worst at or near half
filling and improves as the density decreases from $n=1$.
One cause of this effect is that the size of the Hilbert space is
maximal at half filling and decreases rapidly at large doping.
This effect is magnified in the 16--site system relative to the
32--site system, as seen in
Fig.~\ref{fig4}, because a substantial
proportion of the Hilbert space is retained in the diagonalization
step at large doping.
Another possible cause is the reduction in the effective
strength of the electron--electron scattering as the system is doped
away from half filling.
The effective interaction strength depends on the
ratio of $U$ and the density of states at the
Fermi energy, which becomes smaller with doping, in weak coupling.
As the effective interaction becomes smaller, the electrons become more 
localized in momentum space, which should be favorable for the
convergence of the momentum--space DMRG.
In the real--space approach, Fig.~\ref{fig4} shows that
the error in the ground--state energy first increases as the system is
doped slightly away from half filling, then decreases upon further doping.
As discussed previously, the charge degrees of freedom are localized
in the half--filled insulator, leading to improved convergence for the
real--space algorithm.
For any finite doping, the system immediately becomes metallic,
i.e., some charge degrees of freedom become delocalized, 
leading to a reduction in accuracy.
As the system is doped further from half filling the reduction in the
size of the Hilbert space leads to an improvement in accuracy, as in
the momentum--space approach.

\subsection{Extrapolation to $\bm{m} \rightarrow \infty$}

DMRG calculations have a truncation error which is reduced
by increasing the number $m$ of density--matrix eigenstates
kept.~\cite{steve1,book}
It is important to analyze the scaling of DMRG
results as a function of $m$ to estimate DMRG errors quantitatively.
In real--space DMRG calculations one generally observes that
energy errors $\Delta E(m)$ are proportional to the 
discarded weight $P_m$,~\cite{bonca,jeckel} 
provided that the DMRG has converged to the right target state. 
Here the discarded weight $P_m$ is defined as
the total weight (sum of the density--matrix eigenvalues)
of the discarded density--matrix eigenstates, averaged
over a sweep through all lattice sites
in the finite--system DMRG algorithm.
Thus, it is possible to extrapolate DMRG 
eigenenergies to the limit $P_m \rightarrow 0$. This procedure produces 
extrapolated energies which are closer to 
the exact eigenenergies than the DMRG energies calculated for a 
given value of $m$. 
Moreover, the extrapolation yields reliable quantitative error 
margins for the eigenenergies.

In momentum space, however, we have found that
the linear relationship between
the energy errors $\Delta E(m)$ and 
$P_m$ often does not hold, even for small $P_m$. 
In fact, we find that the dependence of $P_m$ on $m$ can even be
non--monotonic.
An extrapolation to vanishing discarded weight $P_m \rightarrow 0$ is
therefore generally not possible.
That such a non--monotonic behavior is found at all is surprising at
first glance because
the discarded weight $P_m$ of the \textsl{exact} density matrix
for the system decreases monotonically with increasing $m$ per
definition.
An exact density matrix for the ground state
of the Hubbard model can be calculated numerically
in small systems using exact diagonalization.
Using the results of such a calculation on a $N=12$ lattice, we have
found that the density--matrix eigenvalues $w_i, i=1,2,\dots$ 
appear to decrease exponentially as a function of $i$
in the asymptotic regime $i \gg 1$. 
As a consequence, the exact discarded weight $P_m =\sum_{i=1}^{m} w_i$ 
and the corresponding energy error $\Delta E(m)$ must also decrease 
exponentially with increasing $m$.
This is observed for density matrices calculated in
the momentum--space approach as well as those 
obtained in the real--space approach (both for open and periodic
boundary conditions).
Such an exponential falloff of the density--matrix eigenvalues has
been found for exactly solvable models.\cite{peschel}
In an actual DMRG calculation, however, 
the density matrix is calculated self--consistently.
Thus different density matrices can be obtained for different $m$,
and $P_m$ can, in principle, be an arbitrary function of $m$, except
for the condition $\lim_{m \rightarrow \infty} P_m =0$.
We expect such effects to be largest where $P_m$ is large and the
self--consistently determined density matrix is a poor approximation
to the exact one.  

In the momentum--space approach, the error in the DMRG energy 
$\Delta E(m)$ 
does not decrease exponentially with increasing $m$,
but rather
shows a power--law behavior in~$1/m$ in the limit $m \gg 1$.
We therefore extrapolate the momentum--space DMRG results to vanishing
truncation error by performing a least--squares fit of the DMRG
energies $E_{\text{DMRG}}(m)$ 
for several $m$ to a $n$-th order polynomial in $1/m$,
\begin{equation}
E_{\text{fit}}\left (\frac{1}{m} \right ) = E_\infty + \frac{a_1}{m}
+ \frac{a_2}{m^2} + \dots  + \frac{a_n}{m^n}  \; .
\label{eq:polynomial}
\end{equation}
An extrapolated energy for vanishing truncation errors 
($m \rightarrow \infty$) is given directly by the fit parameter
$E_\infty$.
The energy $E_{\text{DMRG}}(m)$ must be a monotonically decreasing
function of $m$ since the DMRG is a variational method and increasing $m$ 
means increasing the variational subspace dimension.
Therefore, $E_{\text{DMRG}}(1/m)$ must satisfy the constraint
$\frac{dE_{\text{DMRG}}(x)}{dx} > 0$ for $x$ in the
range $0 < x \leq 1/m' \leq 1$, where $m'$ is smaller than 
or equal to the smallest value of $m$ used in the fit.
An obvious consequence of this constraint is that the first--order term
in the polynomial $E_{\text{fit}}(1/m)$ must satisfy $a_1 \geq 0$.
We have found that the best fit under this constraint systematically gives
$a_1=0$.

\begin{figure}
\includegraphics[width=8cm]{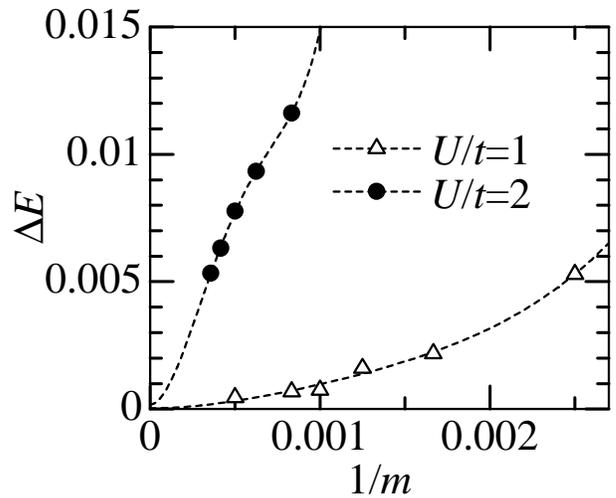}
\caption{DMRG error in the ground--state energy 
as a function of $1/m$ for the one--dimensional Hubbard 
model on a 32--site lattice.  
The lines are least--square fits to a fourth--order polynomial, 
$n=4$ in Eq.~(\protect\ref{eq:polynomial}).  }
\label{fig5} 
\end{figure}

Fig.~\ref{fig5} shows the DMRG  errors $\Delta E$ 
in the ground--state energies of the one--dimensional 32--site Hubbard 
model  for $U/t=1$ and $2$ and the results of least--square fits to 
a fourth--order polynomial, $n=4$ in Eq.~(\ref{eq:polynomial}). 
The DMRG errors $\Delta E(m)$ for the largest value of $m$ used 
are $4.5 \times 10^{-4}$ ($m=2000$)
and $5.3 \times 10^{-3}$ ($m=2800$) for
$U/t=1$ and $U/t=2$, respectively. 
The accuracy is greatly improved by the polynomial fit and the 
$m \rightarrow \infty$ extrapolation.
The errors in the corresponding
extrapolated ground--state energies per site are  
$2.9 \times 10^{-5}$ for $U/t=1$ and $4.5 \times 10^{-5} $ 
for $U/t=2$.  

To illustrate the benefit of extrapolating DMRG energies
to vanishing truncation errors we now consider  
the quasi--particle gap $\Delta_{\text{qp}}$ of the one--dimensional Hubbard
model at half filling. 
The quasi--particle gap is defined by
\begin{equation}
\Delta_{\text{qp}} = E_0(N+1;N) + E_0(N-1;N) - 2 E_0(N;N)   
\; ,
\label{eq:gap}
\end{equation}
where $E_0(N_{\text{e}};N)$ is the ground--state energy of a system
with $N$ sites and $N_{\text{e}}$ electrons with minimal
$S_z$, i.e., $N_\uparrow = N_\downarrow$ or 
$N_\uparrow = N_\downarrow +1$. 
The charge conjugation symmetry of the Hubbard model
implies that $E_0(N+1;N) = E_0(N-1;N) + U$, 
so only one of these two energies need be computed.
Some results for $\Delta_{\text{qp}}$ are shown in Table~\ref{table1}.
If one calculates the quasi--particle gap with the DMRG results for 
$E_0(N_{\text{e}};N)$ obtained for fixed values of $m$, the magnitude
of errors in $\Delta_{\text{qp}}$ fluctuates widely.
The origin of this behavior has two competing sources.
On the one hand, DMRG errors in the eigenenergies
$E_0(N_{\text{e}};N)$ tend to be systematic for similar 
values of $N$ and $N_{\text{e}}$ 
and cancel when calculating Eq.~(\ref{eq:gap}).
Thus, the absolute error in $\Delta_{\text{qp}}$ can be smaller than the 
error
in $E_0(N_{\text{e}};N)$ as seen in Table~\ref{table1} for the case $U=2t$.
On the other hand, the eigenenergies are extensive 
quantities 
[i.e., $E_0(N_{\text{e}};N)$ scales with $N$ 
for constant density $N_{\text{e}}/N$], 
while $\Delta_{\text{qp}}$ is an intensive quantity 
(i.e., $\Delta_{\text{qp}}$
tends to a constant for increasing $N$ and constant density $N_{\text{e}}/N$).
Thus, even small but non--systematic errors in $E_0(N_{\text{e}};N)$ 
immediately result in much larger relative errors in $\Delta_{\text{qp}}$.
As a consequence, the values of $\Delta_{\text{qp}}$ 
(or similar physical quantities) obtained for a given 
number of density--matrix eigenstates kept might be accurate
but there is considerable uncertainty about their accuracy.
The extrapolation of the ground--state energies
$E_0(N_{\text{e}};N)$ to vanishing truncation errors allow us
to eliminate this uncertainty and even to improve the precision
of our results.
For instance, in Table~\ref{table1} one can see that the errors in
$\Delta_{\text{qp}}$ calculated with extrapolated ground--state energies are
up to one order of magnitude smaller than for the largest
value of $m$ used ($m=2000$).

\begin{table*}
\caption{Ground--state energies and quasi--particle gap,
Eq.\ (\protect\ref{eq:gap}),
for the one--dimensional Hubbard model with
$N=32$ sites.}
\label{table1}
\begin{ruledtabular}
\begin{tabular}{ccddd}
$U/t$& &\multicolumn{1}{c}{$m=2000$}& \multicolumn{1}{c}{extrapolated}&
\multicolumn{1}{c}{exact}\\
\colrule
1& $E_0(N;N)/t$& -33.20078 & -33.21423 & -33.21515\\
& $E_0(N+1;N)/t$&  -32.64191 & -32.65757 & -32.65687\\
& $\Delta_{\text{qp}}/t$& 0.12186 & 
0.11472 & 0.11515\\
\colrule
2& $E_0(N;N)/t$& -26.80161 & -27.01970 & -27.01826\\
& $E_0(N+1;N)/t$& -25.61577 & -25.83275 & -25.83170\\
& $\Delta_{\text{qp}}/t$& 0.37737 & 
0.37390 & 0.37311\\
\end{tabular}
\end{ruledtabular}
\end{table*}

\section{Dispersion of spinon excitations}
\label{sec:spinon}

An advantage of the momentum--space method is that
momentum--dependent quantities can be easily calculated.
In this section, 
we investigate the dispersion of the spinon
excitation $E_{\text{s}}(k)$
at half--band filling for both the one--dimensional Hubbard
model with nearest--neighbor hopping and the $1/r$--Hubbard model; 
the spinon spectrum is known exactly in both cases.

The lowest spin--triplet excitation with momentum $k$ in a system of
size $N$ with a singlet ground state is given by
\begin{widetext}
\begin{equation}
\varepsilon_{\text{t}}(k;N)=E_0(N_{\text{e}}/2 +1,N_{\text{e}}/2-1,k+k_0;N)
-E_0(N_{\text{e}}/2,N_{\text{e}}/2,k_0;N) 
\; , 
\label{eq:triplet}
\end{equation}
where $E_0(N_\uparrow,N_\downarrow,k;N)$ 
denotes the energy of the lowest state
with $N_\uparrow$ ($N_\downarrow$)
up-spin (down-spin) electrons and total momentum~$k$, 
and $k_0$ is the momentum of the singlet ground state.
In one--dimensional spin--1/2 systems, a spin--triplet excitation 
is composed of two or more spin--1/2 spinons which are gapless
in the thermodynamic limit.
Therefore, the lowest--energy spin--one excitation allows us to map 
out the spinon dispersion as a function of momentum
\begin{equation}
E_{\text{s}}(k) = \varepsilon_{\text{t}}(k;N\to\infty) \quad ; 
\quad 0\leq k\leq \pi \; . 
\label{eq:spinon}
\end{equation}

\begin{figure*}
\includegraphics[width=7.5cm]{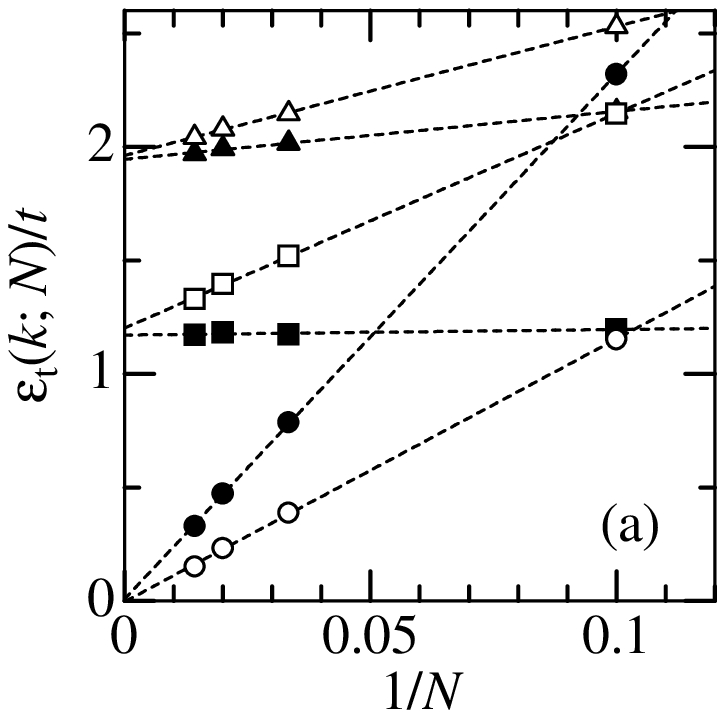}\hspace*{0.5cm}
\includegraphics[width=7.5cm]{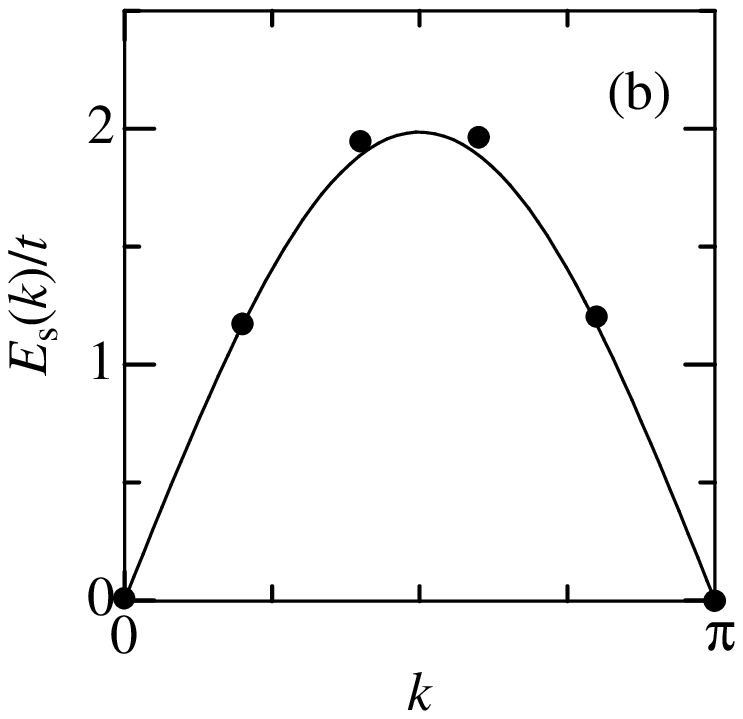}
\caption{Spin excitations in the one--dimensional Hubbard 
model at half filling for $U/t = 0.4$.  
(a) Energy $\varepsilon_{\text{t}}(k;N)$ of the lowest triplet excitation 
calculated with DMRG as a function of the inverse system size $1/N$ 
for momenta $k = 0$ (filled circles), $0.2 \pi$ 
(filled squares), $0.4 \pi$ (filled triangles), $0.6 \pi$ 
(open triangles), $0.8 \pi$ (open squares), and 
$\pi$ (open circles).  
Dashed lines are linear fits in $1/N$.
(b) Spinon dispersion $E_{\text{s}}(k)$ 
in the thermodynamic limit $N \to \infty$. 
The circles show the results obtained by extrapolation of the 
DMRG data in (a) [see Eq.~\ref{eq:spinon}].
The solid line is the exact result obtained in 
Ref.~\protect\onlinecite{takahashi}.}
\label{fig6} 
\end{figure*}

In Fig.~\ref{fig6}(a), we show the lowest triplet excitation
$\varepsilon_{\text{t}}(k;N)$
of the half--filled Hubbard model with nearest--neighbor hopping
at $U/t=0.4$ for several momenta $k$ and system sizes $N$ 
up to 70 sites.
The excitation energies scale approximately linearly 
with~$1/N$ for all momenta but the slope varies considerably
for the different $k$.  
This finite--size scaling is readily understood from the
exact result at $U=0$ for a closed--shell system,
\begin{equation}
\varepsilon_{\text{t}}(k;N)= 2t\sin(k) + 8t \sin\left(\frac{k}{2}\right)
\sin\left(\frac{\pi}{2N}\right)\sin\left(\frac{k}{2}-\frac{\pi}{2N}\right)
\; ,
\end{equation}
where $k=2 \pi n/N$ and $n=1, \ldots, N/2$.
We therefore extrapolate the energies $\varepsilon_{\text{t}}(k;N)$ to the 
thermodynamic limit using a linear fit in $1/N$ 
to obtain the spinon dispersion $E_{\text{s}}(k)$
though Eq.~(\ref{eq:spinon}).  
The results for $E_{\text{s}}(k)$ are shown in
Fig.~\ref{fig6}(b) and compared with the exact spinon
dispersion\cite{takahashi} for $U/t=0.4$.
One sees that our numerical results for $E_{\text{s}}(k)$
agree very well with the analytical curve.  
In particular, $\varepsilon_{\text{t}}(k=0,\pi;N \to \infty)=0$ 
within the accuracy of the extrapolation.  
Note that because of the weak interaction $U=0.4t$ considered 
in this example, DMRG is very accurate and extrapolation of 
eigenenergies as a function of the number $m$ 
of density--matrix states is not necessary. 
Thus, we have used fixed numbers of states $m=400,800,1200,2000$ 
for $N=10,30,50,70$, respectively.
Actually, errors due to the infinite--system extrapolation 
are at least an order of magnitude larger than the DMRG error in 
the ground--state energy (per site), which is $2.7 \times 10^{-4}$
in the worst case, $N=70$.

In the $1/r$--Hubbard model, the spinon spectrum at half band--filling
is given by~\cite{florian}
\begin{equation}
E_{\text{s}} (k)=\frac{1}{4}
\left(\sqrt{W^2+U^2-\frac{4WU (k-\frac{\pi}{2})}{\pi}}
+\frac{2W (k-\frac{\pi}{2})}{\pi}-U \right) 
\quad , \quad 0 \leq k \leq \pi 
\label{Spinone1overrHubbard}
\end{equation}
\end{widetext}
in the thermodynamic limit.
In order to form the lowest spin--triplet excitations, 
two spinon excitations are necessary.
If they have momenta~$k_1=k$ and~$k_2=0$, the spin--triplet
excitation will have total momentum $k$ and 
excitation energy $\varepsilon_{\text{t}}(k)=E_{\text{s}}(k)$ 
with respect to the ground state, as $E_{\text{s}}(k_2=0)=0$.
In Fig.~\ref{fig7}, this analytical result for an infinite system
is compared to our momentum--space DMRG results in finite systems 
of size $N=24$ and $N=32$ for $U/t=1$.
We observe an almost perfect agreement, implying the absence
of significant finite--size effects.  
This is a consequence of the linear dispersion
and weak coupling considered here.   
It can be inferred from the results of Ref.\ \onlinecite{florian}
that the dispersion of spin--triplet excitations
has no explicit dependence on the system size to first order 
in $U/W$
\begin{equation}
\varepsilon_{\text{t}} (k;N) = tk \left( 1 - \frac{U}{W}\right) \; .
\end{equation}
Therefore, the finite--size corrections are of the order
$(1/N)(U/W)^2 \ll 1$ and are negligible in the results
for $U/t=1$ and $N \geq 24$ presented in Fig.~\ref{fig7}.  
As in the nearest--neighbor hopping case, DMRG errors in the energy
are negligibly small compared to the spinon bandwidth because
of the weak interaction used here.
Thus in Fig.~\ref{fig7}
we show DMRG results for a fixed number $m$ of density--matrix states 
($m=800$ for $N=24$ and $m=1600$ for $N=32$)
instead of results extrapolated to the $m\rightarrow \infty$ limit.
These results, along with the results for the one--dimensional Hubbard model
discussed previously,
show that the low--lying spin--excitation spectrum can be accurately
calculated using the momentum--space DMRG, at least in the 
weak--coupling limit.   

\begin{figure}
\includegraphics[width=7.5cm]{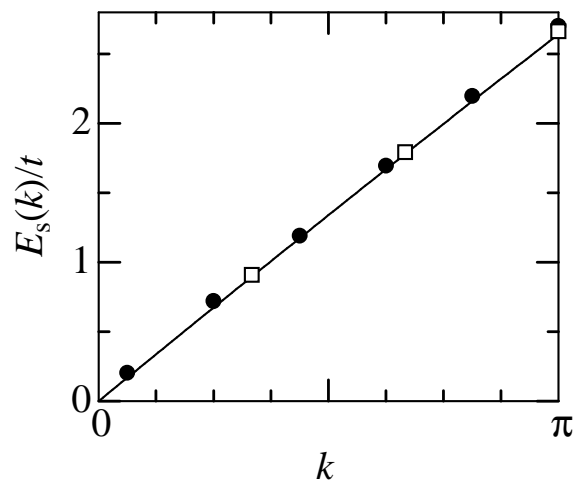}
\caption{Spinon spectrum $E_{\text{s}}(k)$
of the $1/r$--Hubbard model for $U/t = 1$.
DMRG results for finite systems with $N=24$ sites
(squares) and $N=32$ sites (circles).
The solid curve is the exact result for an infinite
system [Eq.~(\ref{Spinone1overrHubbard})].}
\label{fig7} 
\end{figure}

\section{Momentum distribution}
\label{sec:momdens}

Another quantity which is easily accessible 
to the momentum--space DMRG is the single-particle
momentum distribution 
\begin{equation}
n(k)=\frac{1}{2}
\sum_\sigma 
\left\langle n_{k \sigma} \right\rangle \; .
\end{equation}
We have calculated $n(k)$ in the ground state of 
the one--dimensional Hubbard model with nearest--neighbor hopping and 
of the $1/r$--Hubbard model at half band--filling using DMRG.
Since DMRG truncation errors are typically larger for quantities
such as $n(k)$ than for eigenenergies, it is crucial to examine
the effect of varying $m$.
The relative size of the finite--size effects is also important
because one is interested in the behavior of $n(k)$ in the
thermodynamic limit.
In the following, we compare our DMRG results to 
analytic results for~$n(k)$ on infinite lattices in the limit of both
large and small interaction strength. 

\subsection{Nearest--neighbor hopping}
\label{nofkonedim}

In the ground state of the one--dimensional Hubbard model
with nearest--neighbor hopping, the distribution $n(k)$ 
is symmetric, $n(-k) = n(k)$. Therefore, we show results
for $0 \leq k \leq \pi$ only.

We compare our DMRG results
with an approximate expression for $n(k)$ proposed recently by 
Koch and Goedecker.~\cite{koch} 
They make the ansatz that the real--space one--particle density matrix 
$\langle c^{\dag}_{i \sigma} c_{j \sigma} \rangle$ 
of the interacting system 
can be written as a product of
the density matrix for the non--interacting system 
and an exponential decay factor that is a function of the
particle--hole distance.
The corresponding momentum distribution is
\begin{equation}
n_{\text{KG}}(k)=\frac{1}{2}+\frac{1}{\pi}
\arctan \left( \frac{\cos (k)}{\sinh(\gamma)}\right) \; .
\label{weakcouplingKoch}
\end{equation}
Here $1/\gamma$ denotes the decay length and is given by  
\begin{equation}
\frac{e^{-\gamma}}{\pi} =
\left\langle c_{i+1 \sigma}^\dagger c_{i \sigma} \right\rangle
= \int_0^\infty dx \frac{J_0^2(x) - J_1^2(x)}{1 + \exp(Ux/2t)} \; ,
\label{eq:kochgamma}
\end{equation}
where $J_0(x)$ and $J_1(x)$ are Bessel functions of the first
kind.
In Ref.~\onlinecite{koch} it is found that exact diagonalization
calculations on systems of up to 16 sites agree well with this form
for $U/t \agt 6$.
For smaller $U/t$, deviations are seen for wavevectors 
$k\approx k_{\text{F}}$.

\begin{figure}
\includegraphics[width=7.5cm]{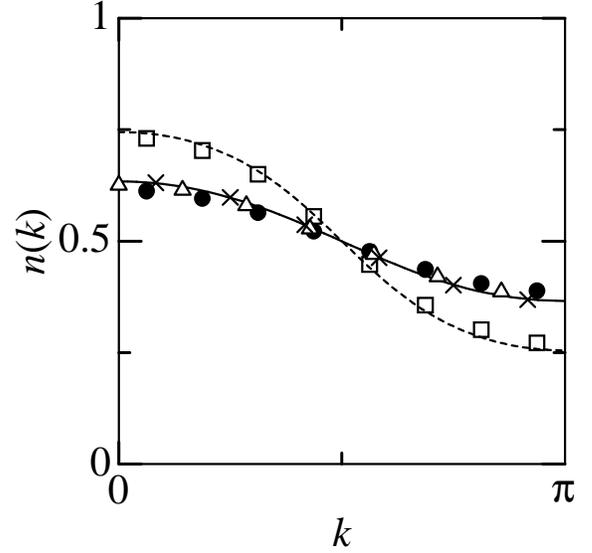}
\caption{Ground state
momentum distribution function $n(k)$ of the one--dimensional 
half--filled Hubbard model at $U/t=20$ 
for system sizes $N=12$ (crosses), $N=14$ (triangles) and 
$N=16$ (circles) keeping $m=2000$ states, and at $U/t=10$ for $N=16$ sites
(squares) keeping $m=1200$ states.  
The lines correspond to the ansatz~(\protect\ref{weakcouplingKoch}).  }
\label{fig8} 
\end{figure}

Expansion of Eq.~(\ref{eq:kochgamma}) in the strong-coupling limit
yields
\begin{equation}
\gamma= -\ln \left ( \frac{2 t \pi \ln 2 }{U} \right )
\end{equation}
and one recovers the perturbative result\cite{takahashi2}
\begin{equation}
n(k) = \frac{1}{2} - \frac{2 \ln(2) \varepsilon(k)}{U} 
\quad ; \quad U \gg t \; .
\label{nkHubbardlargeU}
\end{equation}
In Fig.~\ref{fig8}, we compare Eq.~(\ref{weakcouplingKoch}) 
for $k>0$ with
the momentum--space DMRG results for $U/t=10$ and $U/t=20$
on small finite lattices.
We use periodic boundary conditions ($\phi=0$) for systems
with $N=4n+2$ sites 
and antiperiodic boundary conditions ($\phi=\pi/N$) for 
system with $N=4n$ sites (where $n$ is an integer)
to ensure closed--shell configurations.
The momentum--space DMRG calculations agree
reasonably well with the analytical result, but
we note that the agreement becomes less good with increasing system
size for $U/t=20$. 
An analysis of the behavior for different $m$ shows that our DMRG
results underestimate $n(k)$ for $|k| < k_{\text{F}} = \pi/2 $ and
overestimate $n(k)$ for $|k| > k_{\text{F}}$.
Therefore the deviations seen in Fig.\ {\ref{fig8}} are a consequence
of DMRG errors which become larger for the larger system sizes and we
would expect all results to lie on the analytic curve as in 
Ref.\ \onlinecite{koch} in the $m\to\infty$ limit.

\begin{figure}
\includegraphics[width=8cm]{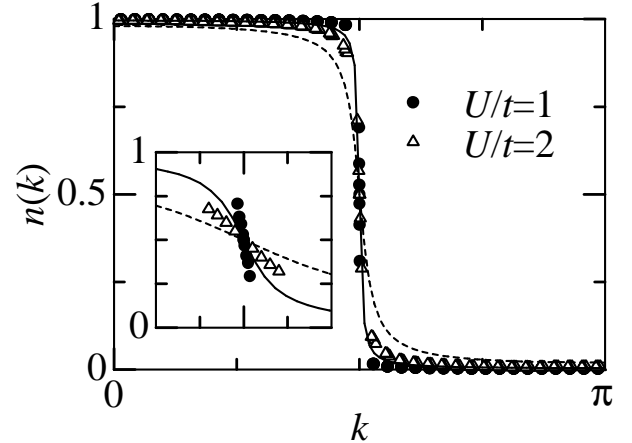}
\caption{Single-particle momentum distribution of 
the one--dimensional half--filled Hubbard model
on a 70--site lattice for different phases $\phi$.  
The circles are for $U/t=1$ (with $m=800$ states kept) and the
triangles for $U/t=2$ (with $m=1200$ states kept).
The lines correspond to Eq.~(\protect\ref{weakcouplingKoch})
for $U/t=1$ (solid) and $U/t=2$ (dashed).
The inset shows a blowup of the region 
$[\pi/2 - 0.0143 \pi,\pi/2 + 0.0143 \pi]$
around the Fermi momentum. }
\label{fig9}   
\end{figure}

In the limit of weak coupling, $\gamma$ is given by~\cite{metzner}
\begin{equation}
\gamma  \approx 7 \zeta (3)\left(\frac{U}{8 \pi t} \right)^2,
\end{equation}
where $\zeta(z)$ is the Riemann Zeta--function 
[$\zeta(3)\approx 1.2$]. 
In Fig.~\ref{fig9} we compare Eq.~(\ref{weakcouplingKoch})
with our DMRG results for $U/t=1$ and $U/t=2$ on a 70--site lattice. 
We superimpose results with different phases $0\leq \phi < 2\pi/N$ 
to obtain better resolution in the vicinity of the Fermi 
momentum~$k_{\text{F}}=\pi/2$.
Note that we can do this only in the insulating phase where the
dependence of the ground state on the phase $\phi$ is negligible for
small $\phi$ because the Drude weight\cite{kohn,shastry-suth} (which
is proportional to
$\partial^2 E_0/\partial \phi^2|_{\phi=0}$) vanishes.
Unlike the limit of strong coupling,
the agreement between our DMRG data 
and Eq.~(\ref{weakcouplingKoch}) is not good near~$k_{\text{F}}$,
as seen in the inset of Fig.~\ref{fig9}.
The first derivative of $n(k)$ at $k= k_{\text{F}}$ is 
$n'(k_{\text{F}}) \approx -64$ at $U/t=1$ and 
$n'(k_{\text{F}}) \approx -10.3$ at
$U/t=2$, whereas Eq.~(\ref{eq:kochgamma}) yields 
$n'_{\text{KG}}(k_{\text{F}}) = -23$ at $U/t=1$ and 
$n'_{\text{KG}}(k_{\text{F}}) = -5.0$ at
$U/t=2$.
Although the ansatz~(\ref{weakcouplingKoch}) correctly
describes the overall shape of~$n(k)$ deep in the insulating regime
($U/t\agt 6$), we find that it does not quantitatively recover the 
$U/t\to 0$ and $|k-k_{\text{F}}|\to 0$ scaling limit.  

\begin{figure}
\includegraphics[width=8cm]{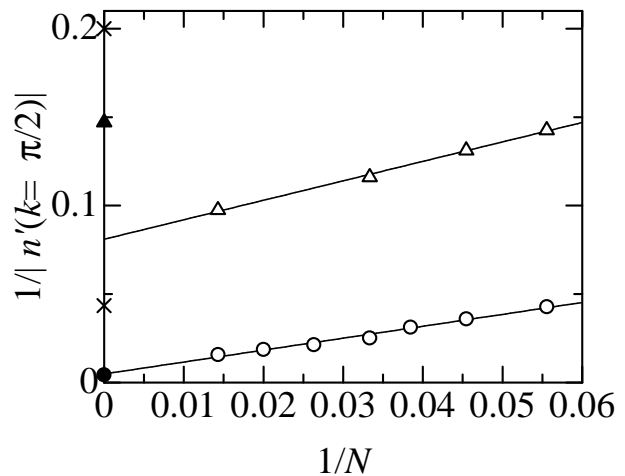}
\caption{
Inverse of the slope at the Fermi wavevector $1/|n'(k = \pi/2)|$ as a
function of inverse system size $1/N$ for $U/t=1$ (open circles) and
$U/t=2$ (open triangles).
The lines are linear fits to the DMRG data.
The filled symbols are the corresponding field--theoretic results and
the crosses are calculated from the Koch--Goedecker ansatz 
(\protect\ref{weakcouplingKoch}).
}
\label{fig10}   
\end{figure}

The behavior of the slope in the scaling limit can be understood by
examining the Green function for the sine--Gordon model.\cite{essler}
The Fourier transform of the equal--time Green function becomes
\cite{esslerprivate}
\begin{equation}
n_{\text{ft}}(k\ge0) = \frac{1}{2} - \frac{Z_0}{\pi} \arctan\left ( 
\frac{v_{\text{h}}(k- \pi/2)}{\Delta_{\text{qp}}/2}
\right ) \; ,
\label{eq:essler}
\end{equation}
where $v_{\text{h}} = 2 t$ is the holon velocity and $Z_0 =0.9219$ has been
calculated in Ref.~\onlinecite{lukyanov}.
The slope at $k=k_{\text{F}}=\pi/2$ is then given by
\begin{equation}
n'_{\text{ft}}(k=k_{\text{F}}) = -\frac{4 Z_0}{\pi \Delta_{\text{qp}}} \; ,
\label{eq:esslerslope}
\end{equation}
which has the values
$n'_{\text{ft}}(k=k_{\text{F}}) = -233$ at $U/t = 1$ and 
$n'_{\text{ft}}(k=k_{\text{F}}) = -6.8$ at $U/t = 2$ when the exact
values for $\Delta_{\text{qp}}$ in the thermodynamic 
limit\cite{ovchin} are used.
In Fig.\ \ref{fig10} we show the scaling with inverse system size of
the DMRG results for $1/|n'(k=k_{\text{F}})|$.
For $U/t = 1$, a $1/N\to 0$ extrapolation (using a linear form) yields
$n'(k=k_{\text{F}}) = -207$, which agrees to within the fit errors with
the field--theoretic result, corroborating both the field--theoretic
and DMRG results in this regime.
For $U/t = 2$, we obtain $n'(k=k_{\text{F}}) = -12$, whose absolute
value is substantially larger (i.e., outside our estimate for the
error) than that of the field--theoretic result.
We believe that this is because $U/t=2$ is large enough so that the
field--theoretic approximation to the momentum distribution of the
Hubbard model begins to show significant deviations from the lattice
result.
This is supported by the fact that the field--theoretic formula
(\ref{eq:esslerslope}) correctly reproduces the strong--coupling form
(\ref{nkHubbardlargeU}), but with a prefactor $4 Z_0/\pi \approx 1.2$
rather than $4 \ln 2 \approx 2.8$,
indicating that the field--theoretic formula underestimates the
slope at larger values of $U$.

\subsection{$\bm{1/r}$--hopping}

In the ground state of the $1/r$--Hubbard model at half filling,
$n(-k) = 1-n(k)$.
The half--filled model describes a metal for $U < W = 2 \pi t$ and an
insulator for $U >W$.

The momentum distribution in the $1/r$--Hubbard mod\-el 
can be calculated in perturbation theory.
For large coupling $U\gg t$, we use 
Takahashi's approach~\cite{takahashi2}.
To this end, we need the spin--spin correlation function
in the Gutzwiller--projected paramagnetic Fermi--sea at half band-filling,
which is the ground state of the Haldane--Shastry spin chain.~\cite{HSmodel}
The spin--spin correlation function for the Haldane--Shastry model
is given by~\cite{florian,GV}
\begin{equation}
\frac{1}{N} \sum_l \langle \Vec{S}_{l} \cdot \Vec{S}_{l+r} \rangle
= (-1)^r \frac{3{\text{Si}}(\pi r)}{4\pi r} \quad , \quad (r\neq 0) \; , 
\end{equation}
where ${\text{Si}}(x)$ denotes the sine-integral. 
The momentum distribution for large $U/t$ is then
\begin{eqnarray}
n(k)&=& \frac{1}{2} -\frac{1}{U}\sum_{r=1}^{\infty}
(-1)^r \frac{\sin(kr)}{r} \left( \frac{3{\text{Si}}(\pi r)}{\pi r} -1\right)
\nonumber \\[3pt]
&=& \frac{1}{2}\left[1-\frac{k}{U}
\left(1-3 \ln \left|\frac{k}{\pi}\right|\right)\right] \; .
\label{nkflorianstrong}
\end{eqnarray}
Note that although the momentum distribution is continuous at 
$k_{\text{F}}=0$, we can expect to observe a large
apparent jump in numerical simulations of finite systems even at
large~$U/t$ because of the sizable logarithmic term
in Eq.~(\ref{nkflorianstrong}).

For small couplings,
standard perturbation theory in~$U/t$ is applicable because the model
reduces to a pure $g_4$~model in the conformal limit.~\cite{florian}
It turns out that the Gutzwiller wave function becomes exact 
in the small--coupling limit so that the momentum distribution for
$U\ll t$ becomes\cite{MV}
\begin{equation}
n(k) =
\left\{
  \begin{array}{clc}
 1-(U^2/W^2) f(k) & \mbox{for} & -\pi < k < 0 \\
 (U^2/W^2) f(k) & \mbox{for} &  0 < k < \pi \\
  \end{array}
\right.
\label{nkflorianweak}
\end{equation}
with $f(k)=3/16-(1/4-|k|/(2 \pi))^2$.

\begin{figure*}
\includegraphics[width=8cm]{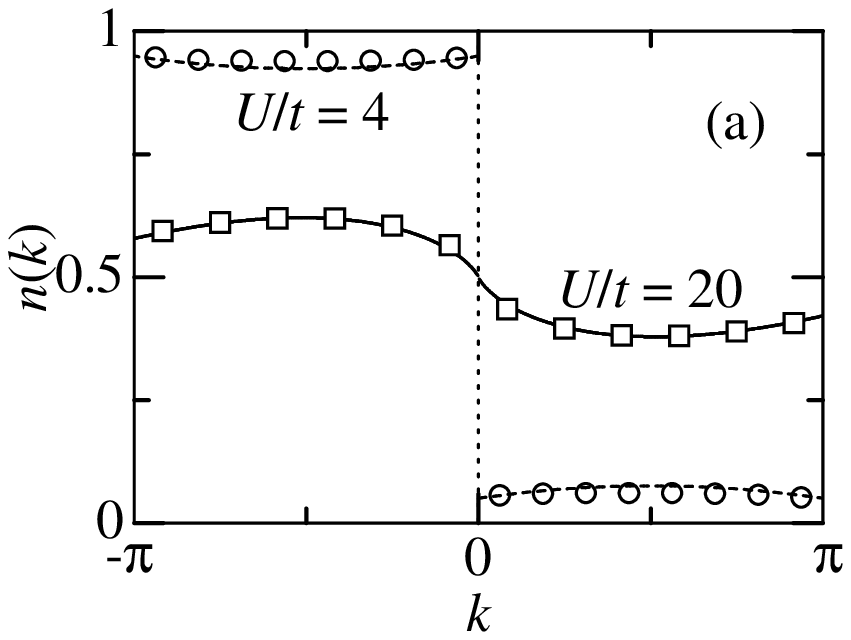}\hspace*{0.5cm}
\includegraphics[width=7.2cm]{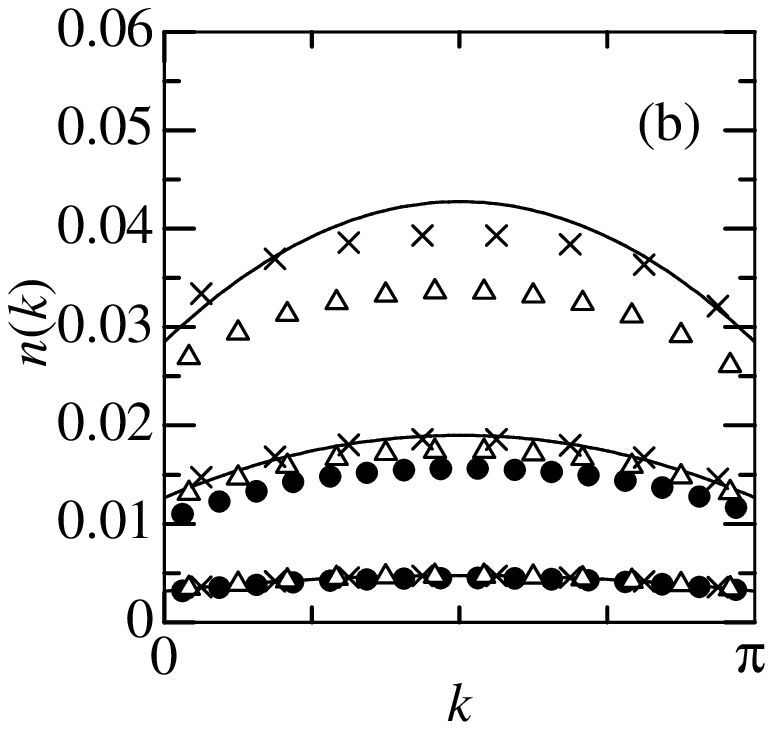}
\caption{Momentum distribution function for the $1/r$--Hubbard model at 
half filling with $m=2000$ states kept.
(a) DMRG results on an $N=16$ lattice for $U/t=4$ and an $N=12$ lattice for
$U/t=20$.
The solid and dashed lines are from the perturbative first--order 
result in~$t/U$, Eq.\ (\protect\ref{nkflorianstrong}), 
and the second--order result in $U/t$, Eq.\ (\protect\ref{nkflorianweak}), 
respectively.  
(b) The weak--coupling regime for $U/t=1, 2, 3$ 
(from bottom to top).  The crosses, 
open triangles and filled circles denote $N=16, 24, 32$.
The solid line results from Eq.\ (\protect\ref{nkflorianweak}).  }
\label{fig11} 
\end{figure*}

Figure~\ref{fig11}(a) displays the momentum distribution function 
for the half--filled $1/r$--Hubbard model 
for $U/t=20$ on an $N=12$ lattice compared with the 
strong--coupling result~(\ref{nkflorianstrong}),
and for $U/t=4$ on an $N=16$ lattice compared with the 
the weak--coupling result~(\ref{nkflorianweak}).
The agreement is very good because
the DMRG errors are negligible for such small systems and because
the finite--size effects are small, at least on the scale of the
figure.

The finite--size effects in weak coupling are more visible in
Fig.~\ref{fig11}(b).
One can see that the perturbation theory, Eq.\ (\ref{nkflorianweak}),
agrees well with the numerical results at all system sizes for $U/t=1$.
However, deviations that become larger with system size can be seen
for $U/t=2$.
The results for the smaller systems incorrectly suggest that
Eq.\ (\ref{nkflorianweak})~applies perfectly and that finite--size effects
are absent. 
This effect is even more marked at $U/t=3$
where the analytic weak--$U$ result agrees almost perfectly with
the numerical data for $N=16$, but
the data for $N=24$
reveal that the agreement would be worse in the thermodynamic limit.
Apparently, the finite--size effects are approximately the same order
and sign as the higher--order corrections in $U/t$ here.

\section{Discussion and outlook}
\label{sec:outlook}

In this work, we have examined in detail the application of the
Density Matrix Renormalization Group to the momentum--space
representation of the Hubbard model.
We have treated three different dispersion relations corresponding to
the one--dimensional chain with nearest--neighbor hopping, the
one--dimensional chain with hopping that decays as $1/r$, and the
two--dimensional square lattice with isotropic nearest--neighbor
hopping.
While the one-- and two--dimensional nearest--neighbor hopping cases
were treated previously by Xiang,\cite{xiang} here we 
have extended the scope of his results and have addressed some
issues raised by his work.
In particular, we have taken advantage of the Bethe Ansatz exact
solution,\cite{liebwu} which yields the exact ground--state energy to
within machine precision for all system sizes, to make more extensive
studies of the convergence for the one--dimensional model with the
number of density--matrix eigenstates kept, $m$.
We have examined the effect of system size,
interaction strength and band filling at up to $m=4000$.

For all parameters and models, we have found systematic variational
convergence with $m$ to the true ground state; this does not seem to
clearly occur in Xiang's, at least in his $U/t=4$ results for the
one--dimensional system.
However, the convergence seems to be
slower than exponential for the range of $m$ accessible to us.
The accuracy decreases regularly with system size when the other
parameters are fixed, 
as one finds for the real--space DMRG.
The accuracy also becomes rapidly worse with interaction $U$, as
also found by Xiang.
Our more extensive set of interaction strengths  
indicates that the behavior is quite regular as a function of
$U$; convergence does not break down at a particular $U$ value.

In the one--dimensional Hubbard model with periodic boundary
conditions, a comparison with the real--space DMRG applied to the same
system with the same $m$ indicates that the accuracy of the
real--space method is better for $U/t \agt 1$.
The dependence of the accuracy on band filling is
weak in momentum space, except on small system sizes for which the
proportion of the Hilbert space kept changes
drastically with filling.

For the two--dimensional system, we have restricted ourselves to the
half--filled $4\times4$ system, the largest for which exact
diagonalization data are available, and have compared with the
real--space DMRG.
We find that the momentum--space method is more accurate than the
real--space method (for the same $m$) when $U \alt 8 t = W$.

One crucial issue raised by Xiang is the dependence of the accuracy
on dimensionality or, relatedly, the range of the hopping.
He speculated that such effects would be smaller for the
momentum--space DMRG than for the real--space DMRG, a speculation that
we have confirmed here.
We emphasize however that the choice of values of interaction strength
which are compared is important.
It is our opinion that a reasonable choice is the interaction divided
by the bandwidth, $U/W$ (at identical filling and number of lattice
sites).
The bandwidth sets the energy scale for many physical phenomena and
also is relevant to the strength of the coupling in perturbation
theory.
(A possibly useful alternative might be the ratio of the interaction
strength and the density of states at the Fermi energy, the coupling
parameter in weak--coupling perturbation theory.)
For given $U/W$ (as opposed to given $U/t$), the accuracy of the
momentum--space DMRG is lower in two dimensions than in one, an effect
which becomes smaller as the interaction becomes larger.
In one dimension, changing to the longer--range $1/r$ hopping has
little effect on the accuracy at weak $U/t$, although the accuracy
does become somewhat worse as $U/t$ is increased.
We therefore conclude that while the performance of the
momentum--space DMRG is less dependent on range of the hopping or
dimensionality than the real--space DMRG, there is still some
effect.

While the ground--state energy is an important indicator of
convergence, it is not a particularly useful quantity in determining
the physical behavior of a system.
We have therefore examined a number of other quantities that are
easily accessible to the momentum--space DMRG, which also yield useful
physical information.
Gaps formed from differences in energies provide important information
about the excitation spectrum.
We have examined the single--particle gap as well as the
momentum--dependent triplet gap for both one--dimensional models at
half filling.
For the single--particle gap, extrapolation in $1/m$ is crucial in
obtaining consistent accuracy because of cancellation of variational
errors and because the gap is an intensive quantity obtained by
subtracting extensive energies.
We have found that a direct extrapolation using a polynomial in $1/m$
is the best method because the error in the ground--state energy is
not well correlated with the weight of the discarded density--matrix
eigenvalues, unlike in the real--space DMRG.
For the spin excitation spectrum, we have treated parameter values
for which extrapolation in $m$ is unnecessary and found that the
finite--size effects are substantial for the
nearest--neighbor--hopping chain.
For the $1/r$--Hubbard model, finite--size effects were quite small.
In both cases, the size--extrapolated spectrum agrees well with exact
results.

Finally, we have examined the momentum distribution function.
For the nearest--neighbor chain, we compare with an analytical
ansatz of Koch and Goedecker.\cite{koch}
At strong coupling, we find very good agreement aside from deviations
due inaccuracy of the DMRG results.
At weak interaction, the DMRG results agree with field--theoretic
results,\cite{essler} whereas there is significant deviation from the
Koch--Goedecker ansatz.
For the $1/r$--Hubbard model, agreement with weak and strong coupling results
is good, although finite--size corrections with the same 
sign as higher order terms in $U/t$ provide better
agreement for small system sizes than is justified in the
thermodynamic limit.

In summary, the momentum--space DMRG can be a useful tool for the
Hubbard model at weak to intermediate coupling.
While it is competitive with the real--space DMRG only at quite weak
coupling for the one--dimensional model (even with periodic boundary
conditions), it is competitive up to significantly stronger coupling
for longer--range hopping or in two dimensions.
It should be noted that the accuracy of the momentum--space DMRG is
generally significantly lower than that of the more favorable cases
for the real--space DMRG, and can be considered a
``numerically exact'' method only with reservations.
It can, however, be a useful variational method where no more exact
methods are available if its limitations are well understood.
The ease of calculation of momentum--dependent quantities is very
useful -- such quantities are often not available on large systems
even for well--understood models.
For these quantities, care must be taken with respect to accuracy and
finite--size effects, but we have found them to be
well--behaved within these limitations.

The momentum--space DMRG code used here is far from maximally optimized.
With similar optimizations as used in the real--space program such as
the use of wavefunction transformations to improve the initial guess for
the target state in the diagonalization,\cite{steve3}
it should be possible to keep significantly more density--matrix
eigenstates for given numerical effort in the momentum--space method
than in the real--space method applied to the same system.
Use of such a better optimized program might increase the range of
applicability of the momentum--space DMRG somewhat.
In addition, since the Hubbard model has a local interaction in
real--space, the interaction is quite non--local in momentum space.
The momentum--space DMRG could quite possibly be better suited to
models with longer range interaction in real space, corresponding to
more local interactions in momentum space.
Such directions would certainly be worth exploring in future work.

\begin{acknowledgments}
This work was supported by the Deutsche Forschungsgemeinschaft 
under grant number GE 746/6-1.
We acknowledge helpful discussions with F.H.L\ Essler, S.R.\ White and
T.\ Xiang.
\end{acknowledgments}

\end{document}